\documentclass[aps,prl,twocolumn,groupedaddress]{revtex4-1}
\usepackage{graphicx}
\usepackage[dvips]{color}
\usepackage{amsmath}
\usepackage{amssymb}
\usepackage{amsfonts}

\graphicspath{{./},{Figures/}}

\begin{document}

\title{Jet impact on a soap film}

\author{Geoffroy Kirstetter}
\affiliation{Laboratoire de Physique de la Mati\`ere Condens\'ee, CNRS UMR 7336, Universit\'e de Nice Sophia-Antipolis, 06108 Nice, France}
\author{Christophe Raufaste}
\email[corresponding author : ]{Christophe.Raufaste@unice.fr}
\affiliation{Laboratoire de Physique de la Mati\`ere Condens\'ee, CNRS UMR 7336, Universit\'e de Nice Sophia-Antipolis, 06108 Nice, France}
\author{Franck Celestini}
\email[corresponding author : ]{Franck.Celestini@unice.fr}
\affiliation{Laboratoire de Physique de la Mati\`ere Condens\'ee, CNRS UMR 7336, Universit\'e de Nice Sophia-Antipolis, 06108 Nice, France}

\date{\today}

\begin{abstract}

We experimentally investigate the impact of a liquid jet on a soap film. We observe that the jet never breaks the film and that two qualitatively different steady regimes may occur. The first one is a refraction-like behavior obtained at small incidence angles when the jet crosses the film and is deflected by the film-jet interaction. For larger incidence angles, the jet is absorbed by the film, giving rise to a new class of flow in which the jet undulates along the film with a characteristic wavelength. Besides its fundamental interest, this study presents a new way to guide a micro-metric flow of liquid in the inertial regime and to probe foam stability submitted to violent perturbations at the soap film scale.

\end{abstract}

\pacs{47.55.-t, 47.15.-x, 68.03.-g}
%47.15.-x: laminar flows
%47.55.-t: Multiphase and stratified flows
%68.03.-g: Gas-liquid and vacuum-liquid interfaces
%\keywords{}

\maketitle

\section{Introduction}

Control and manipulation of laminar jets are of paramount importance in the context of miniaturization and use of microfluidic systems. Systems such as inkjet \cite{Badie1997,Dong2006,Brown2010}, encapsulation for biological applications \cite{Loscertales2002,Funakoshi2007}, fiber spinning \cite{GananCalvo2004} rely on the stability of the micro-jet or conversely on its destabilization through the control of the liquid jet atomization or drop-on-demand process. But if technologies such as electro-spray devices \cite{Badie1997,Dong2006,Brown2010}, focused surface vibrations \cite{Yeo2009,Tan2009}  combined or not with flow-focusing techniques \cite{GananCalvo1998,GananCalvo2004} can control the transition between jetting and dripping,  no reliable technique is available to guide a micro-jet inside a medium as simple as air. Recently, rebound on a hydrophobic surface was found to deflect a jet \cite{Celestini2010}, but this process is prevented in most cases by the  spreading of the liquid on the substrate.

Furthermore, control of liquid foam stability is a prerequisite in numerous industrial applications like fire fighting, oil recovery, ore extraction, explosion safety and  food or cosmetics processing \cite{Aubert1986, Weaire1999,Cantat2010}. Liquid foams are made of gas bubbles separated by liquid soap films. Their stability under mechanical solicitations is a major issue: as liquid fraction and soap film thickness are directly related to the osmotic pressure inside the liquid films \cite{Hohler2008}, all kind of mechanical effects which can balanced this pressure can dramatically alter the foam properties through soap film bursting and bubble coalescence.
Violent mechanical perturbations, such as impacts, have recently raised some interest and uses for sound absorption or bomb explosion safety  \cite{Clarck1986}.
Solid particles \cite{LeGoff2008} or liquid drops \cite{Gilet2009} impacting a soap film lose kinetics energy and exhibit a rich variety of behaviors amongst film crossing, bouncing, partial coalescence and formation of satellite droplets. To our knowledge, nothing is known about the soap film stability after the impact of a liquid jet. Conversely to the studies cited above, the soap film is probed by continuous mass and momentum inputs provided by the liquid jet.
 
\begin{figure}[!h]
\includegraphics[width=10cm]{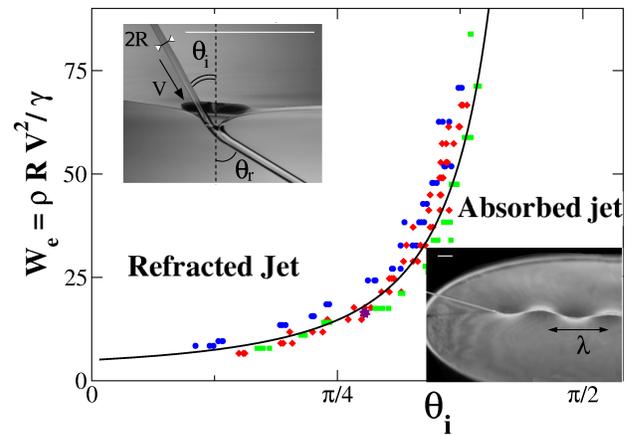}
\caption{\label{diagend} (Color online)
Impact diagram displaying the refraction and absorption regions. Each point corresponds to the onset of the transition from the refraction to the absorption regime obtained for jet radii $R_i = 140$, $200$ and $270$ $\mu$m and the Dreft solution, and for jet radii $R_i = 140$ $\mu$m and the TTAB solution (circles, diamonds, squares and stars respectively). The solid line represents the model detailed in the main text. In the representative pictures, white scale bars  have a length of $5$ mm.
}. 
\end{figure} 
 
We investigate the impact of a liquid jet on a soap film. 
By tuning the jet velocity and/or  incident angle,  two qualitatively different steady regimes are observed.  In the first one, the jet crosses the film without breaking it and is deflected by the film-jet interaction. This feature can be used to guide and control the jet direction. In the second one, a new class of flow is reported: the jet is merged with the film and undulates inside the latter with a characteristic wavelength. A transient state  corresponding to a bouncing jet on the film is also observed. Both regimes are well described using the Weber number ($We = \rho V_i^2 R_i/\gamma$) quantifying the relative importance of inertia and capillarity.  
Simple models are finally  successfully proposed to quantitatively describe both regimes.

\section{Materials and methods}

 We experimentally study  the impact of a laminar liquid jet onto a film of the same composition. We use two solutions: most of the experiments were performed with a soap solution obtained by adding 5\% of commercial dish-washing liquid (Dreft, Procter \& Gamble) to deionised water. To test the robustness of the results, some points were confirmed using a TTAB solution (3g/L).
The experimental set-up has already been described in details in a former study \cite{Celestini2010}. A pressurized chamber is built to inject the liquid at a controlled constant flow rate through a sub-millimetre nozzle, a laminar jet forms at the exit of the latter.
 The incident jet is characterized by its incident angle $\theta_i$, velocity $V_i$ and radius $R_i$. The jet velocity  varies within the range of 1 to 4 m.s$^{-1}$, and several jet radii between $R_i=80$ and $270$ $\mu$m have been used. The injector is placed just above an horizontal soap film maintained by a circular frame, of 10 cm in diameter. 
 We note $\gamma$  the surface tension (equals to $26.2 \pm 0.2$ mN.m$^{-1}$ for the Dreft solution,  $38 \pm 1$ mN.m$^{-1}$ for the TTAB solution) and $\rho$ its density (equals to $10^{3}$kg.m$^{-3}$ in both cases).
 Within our experimental parameters range, based on the jet characteristics, the Reynolds number is always significantly larger than unity.

\section{Results}

Regardless its velocity, radius and incident angle, the jet never breaks the soap film.
Several works have explored the stability of soap films under the impacts of particles \cite{LeGoff2008} or liquid drops  \cite{Gilet2009}. In all regimes explored, the films close after the crossing of the impacting projectiles. 
Pinch-off of the films while they are stretched is found to be the healing mechanism which ensures their continuity as a function of time  \cite{LeGoff2008}. In the present case of impacting jets, such mechanism is not involved in the film stability. The film/jet contact is never broken and the film does not need to close.  Depending on the jet characteristics, two qualitatively different regimes are observed. 
By analogy with optics, we called the first one "refraction-like regime" : the jet crosses the film and is deflected. The second regime is called "absorption" : beyond a critical angle, the jet is trapped by the film and undulates along it. For a given set of input parameters, i.e. the values of $V_i$ and $R_i$, the angle $\theta_i$ for which the transition occurs is recorded. The impact diagram in the ($W_e ,\theta_i $) space of the system  is represented in Fig. \ref{diagend}. The data have been recorded  for the three different jet radii considered in this study. We can see that using the Weber number, the influence of both $R_i$ and $V_i$ is well captured and  that all points collapse onto the same master curve. This scaling therefore demonstrate that the transition between the two regimes is governed by the interplay between capillarity and inertia and that the dissipation inside the jet/film contact zone can be neglected. 
 
\begin{figure}
\includegraphics[width=8cm]{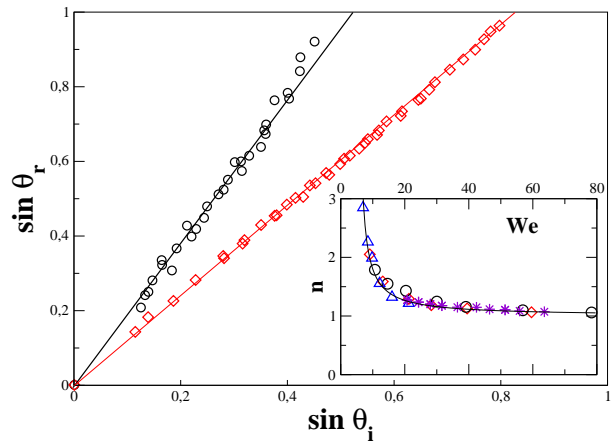}%
\caption{\label{transmission} (Color  online)
$\sin(\theta_r)$ versus $\sin(\theta_i)$ for a radius of $140$ $\mu$m and two incident velocities $V_i=1.3$ and $2.3$ m.s$^{-1}$ (circles and diamonds respectively). Full lines are best linear fits used to calculate the refraction index $n$. Inset :  $n$ versus the Weber number for different velocities, three different jet radii and two solutions: $R_i=80$, 120, and 140 $\mu$m for the Dreft solution (triangles, circles and diamonds  respectively), $R_i=140$ $\mu$m for the TTAB solution (stars). The solid line corresponds to the model described in the main text.}
\end{figure}
 
 \subsection{Refraction regime}
 
We first describe the refraction-like regime appearing at high $We$ numbers or small $\theta_i$ values.  As illustrated in the inset of Fig. \ref{diagend} , the jet is refracted with an angle $\theta_r$. At  high jet velocities and low incidences, almost no visual change
 of the jet and film geometries is observed : $\theta_r$ is almost equal to $\theta_i$ and the film is slightly deformed. As the velocity decreases the influence of the film induces measurable changes in the radius, angle and velocity of the refracted jet.  The sinus of refracted angle is represented as a function of the incident one in Fig. \ref{transmission} for a jet of radius equals to $140$ $\mu$m and two different velocities ($V_i=1.3$ and $V_i=2.3$ m.s$^{-1}$). We observe that the jet is deflected towards the film  and that the lower  the velocity, the higher the deflection. A linear regime can be furthermore identified.  By analogy with optics and Snell-Descartes 's law of refraction, an index $n$ can thus be defined as $ n= \sin(\theta_r) / \sin(\theta_i)$ to quantify the deflection. 
We plot in the inset of Fig. \ref{transmission} the values of $n$ obtained for three different radii ($R_i=140, 200,$ and $270$ $\mu$m) and velocities as a function of $We$. Once again the Weber number is found to be the relevant parameter to rescale all $n$ values on a same master curve. 
We therefore demonstrate that inertia and capillarity are the relevant effects and give an explanation to the counter-intuitive observation that the higher the velocity, the lower the changes. As emphasized by the increase of $n$ as $We$ decreases, capillary forces between the jet and the film influence their respective shape and geometry, while friction inside the jet/film interaction zone, which should be an increasing function of the velocity, does never contribute significantly to the interaction force.

Actually, if dissipation does not play a direct role on the jet-film interaction, its presence is necessary to drift the contact line streamward and imposed a dynamical wetting condition as emphasized in \cite{Raufaste2012}. The contact angle evolves from $90^\circ$ to a value close to $0^\circ$. This is of paramount importance to account for a non-zero interaction force and the jet deflection observed at small $We$.\\

\begin{figure}
\includegraphics[width=7cm]{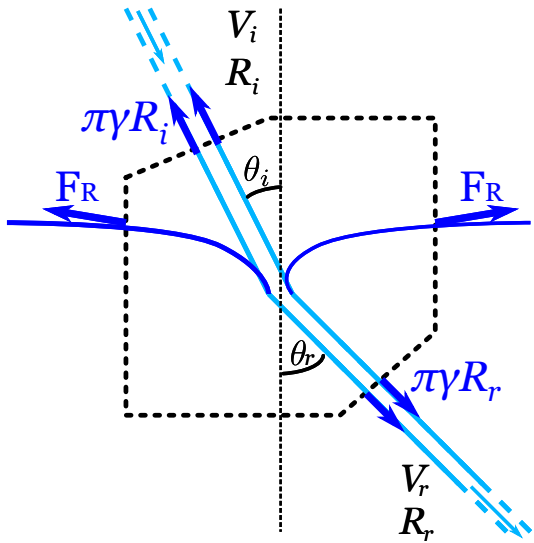}
\caption{\label{Fig:RefractionCloseSystemGeometry} (Color online)
Side view representation of the refraction. Mass and momentum balance equations are performed on an open system enclosing the jet-film interaction zone.}
\end{figure}

A model is proposed to account for the jet-film interaction inside the refraction regime.
Three equations are needed here to account for the mass and momentum  equations and are applied on an open system as the one depicted on Fig. \ref{Fig:RefractionCloseSystemGeometry}.
In what follows, $D$ accounts for the jet flow rate and $F_R$ for the interaction force  inside the refraction regime.  $R_r$ and $V_r$  account for the radius and velocity of the refracted jet.

Assuming  a plug flow inside both the incident and refracted jets, mass balance writes
$$
D   =  \pi R_i^2 V_i  =  \pi R_r^2 V_r 
$$
Momentum balance ($x$ and $y$ projections) is expressed below by balancing the momentum rate changes (left-hand side) and the forces applied on the system (right-hand side):
$$
\begin{array}{ccc}
D(\rho V_r \sin{\theta_r} - \rho V_i \sin{\theta_i})  & = & \pi \gamma (R_r \sin(\theta_r) - R_i \sin(\theta_i)) \\
D(\rho V_r \cos{\theta_r} - \rho V_i \cos{\theta_i})  & = & \pi \gamma (R_r \cos(\theta_r) - R_i \cos(\theta_i))\\
 & & - F_R 
\end{array}
$$
Forces in the momentum balance equations have two contributions: the interaction force $F_R$ assumed perpendicular to the soap film \cite{Raufaste2012} and contact forces (both surface tension and pressure) at the system-jet boundaries (generalization of $\vec{F_2}$ for any refracted angle as described in the absorption regime section).\\
The system of equations can be transformed to expressed the refracted angle $\theta_r$ as a function of the impact parameters and of the interaction force only:
\begin{equation}
(We-1) \sin(\theta_r - \theta_i) = \frac{F_R}{\pi \gamma R_i} \sin(\theta_r)
\end{equation}
Assuming the small inclination limit, we can simplify the system following:
$$
\begin{array}{ccc}
 \sin{\theta_i} &\sim& \theta_i       \\
 \sin{\theta_r} &\sim& n \theta_i      \\
 F_R &\sim&4 \pi \gamma R_i
\end{array}
$$
The last expression assumed a total wetting condition \cite{Raufaste2012} and that the jet radius at the jet-film contact is taken as $R_i$. Experimentally, this radius is found between $R_i$ and $R_r$, and further analysis would be needed to describe the exact contact radius. But  as seen below, such refinement is not necessary to account for the effect and is of second importance.

Finally,  the model leads to
\begin{equation}
n = \frac{We-1}{We-5}
\end{equation}
This expression describes the experimental measurements rather satisfactory (Fig. \ref{transmission}).

\subsection{Transition}

Physically, we might expect the transition to occur for $\sin(\theta_r) = 1$. Given our model, this leads to $\sin(\theta_i) = 1/n$ at the transition or 
\begin{equation}
\theta_i = \arcsin\left( \frac{We-5}{We-1}\right) .
\end{equation}
Again, the agreement is rather satisfactory (Fig. \ref{diagend}) to describe the transition from the refraction regime to the absorption regime.
None the less,  the transition from the absorption regime to the refraction regime can not be described by the same formula, emphasizing an hysteresis behavior. This is mainly due to a new contact zone geometry inside the refraction regime, not accounted for in the model above. We could observe experimentally that this transition occurs for higher velocities/smaller angles, but is difficult to quantify and less reproducible given the fact that the film is strongly deformed, oscillates and is very sensitive to changes close to this transition.

\subsection{Absorption regime}

\begin{figure}
\includegraphics[width=8cm]{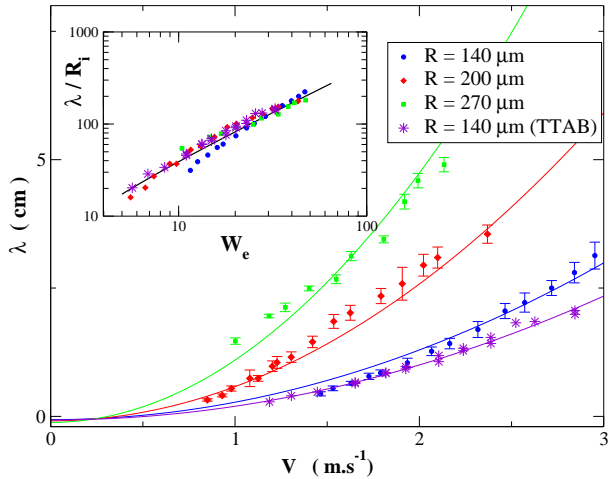}%
\caption{\label{lambda} (Color online)   
Wavelength $\lambda$ versus incident velocity $V_i$ for different radii  $R_i = 140$ , 200  and 270 $\mu$m and the Dreft solution, and $R_i = 140$ $\mu$m for the TTAB solution (circles, diamonds, squares and stars respectively). The solid lines correspond to the model with $\tilde{f}=0.80$. Inset: same set of data plotted under the dimensionless form. }
\end{figure}

We now describe the second regime observed at small $We$ and large $\theta_i$.  In that case, capillary forces are strong enough to compensate the normal component of the jet momentum and we therefore refer to this regime as an``absorption''.
As depicted in Fig. \ref{diagend} the jet follows a wavy trajectory inside the film characterized by its wavelength $\lambda$. The undulation persists over several wavelengths before some relaxation processes merge the jet and the film together and dissipate the kinetics energy continuously provided by the jet.
When beginning the experiment from a refraction situation and by increasing the incident angle (or decreasing the velocity), the system transits to the absorption regime.  
If the jet impacts the soap film with parameters corresponding to the absorption region, a transient stage characterized by a ``reflection''  on the film (i.e. a rebound of the jet on the film) is observed before the absorption occurs. This behavior will be discussed below.  We represent in Fig. \ref{lambda}. the measured value of $\lambda$ as a function of the velocity for three different jet radii. In these experiments the incident angle is fixed to $70$ degrees. The values obtained for $\lambda$ are averaged over two or three different undulations wavelengths (we have checked that the value of the wavelength does not depend on its distance to the impact point). One can clearly see that the higher the velocity, the higher the wavelength (for a fixed jet radius) and that the higher the radius, the higher the wavelength (for a fixed jet velocity).\\

\begin{figure}
\includegraphics[width=8cm]{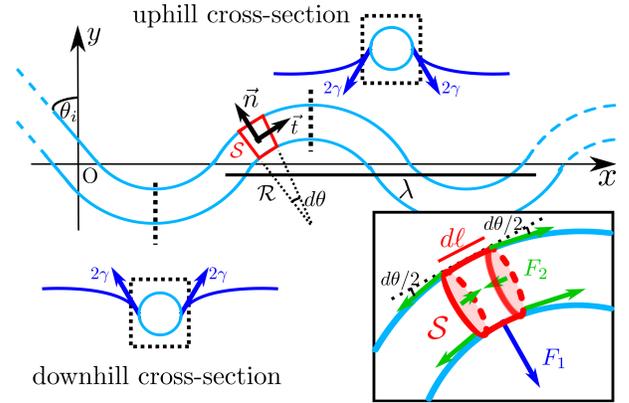}
\caption{\label{Fig:SnakeGeometry} (Color online)
Side view representation of the undulating jet (light blue) inside the film (dark blue). Two cross-sections are displayed showing a reasonable shape of  the film-jet contact. Momentum balance equation is performed on a closed system $\mathcal{S}$. Insert: zoom over the system. }
\end{figure}

To understand this behavior a model is derived to describe the jet properties inside the absorption regime. The second law of Newton is applied to a given portion of fluid as it is depicted in Fig. \ref{Fig:SnakeGeometry}. The length $d\ell$ of this system ${\mathcal{S}}$ is chosen to be small enough compared to the other relevant lengths of the system. In that case, the system has a constant cross-section, which scales as $\pi R_i^2$,  and that is slightly twisted with a radius of curvature ${\mathcal{R}}$.
As will be seen below, the net force exerted on the system is always oriented along the centripetal component of the acceleration, meaning that kinetics energy is constant for the system and consequently its velocity amplitude. By conservation of the flow rate, the radius of the jet is taken constant as well, equals to $R_i$. Consequently the acceleration of the system writes $\vec{a}= V_i^2 / \mathcal{R} $  $\vec{n}$, where $\mathcal{R} = \epsilon |\mathcal{R}|$ is the algebraic radius of curvature, which sign $\epsilon$ holds for the local convexity of the trajectory. 

Assuming that gravity and all sources of dissipation can be neglected, the net force applied on the system has two contributions.
First, the force $\vec{F_{1}}$ applied by the film onto the system $\mathcal{S}$. This force accounts for the deformation of the film due to the film-jet contact. From reasonable film-jet profile transition zone (see cross-section in Fig. \ref{Fig:SnakeGeometry}), the film pulls normally onto the system with a capillary force ranging from 0 to  $4\gamma$ per unit length depending on the geometrical orientation of the film-jet triple line. The average value  $2 \gamma$ per unit length is chosen as the order of magnitude. Without loss of generality, $\vec{F_{1}} = 2\gamma \tilde{f}  d\ell \epsilon \vec{n}$, where $\tilde{f}$ is a constant ranging between 0 and 2.

The second contribution comes from the contact forces applied by the rest of the jet onto the system. The two terms account for the surface tension and pressure forces respectively. For both of them, contributions are found at both the leading and trailing extremities of the system. Amplitude are the same: $2 \pi R_i \gamma -  \pi R_i^2 P= \pi R_i \gamma$, since the pressure $P$ equals the capillary pressure $\gamma/R_i$. Their tangential components compensate each other, but not the normal ones as soon as ${\mathcal{R}}$ is finite. This leads to  $\vec{F_{2}} = 2\cdot \pi R_i \gamma  d\theta/2 \cdot  \epsilon \vec{n}$. 
The momentum balance equations leads to
$$
\rho d\ell \pi R_i^2 V_i^2 / \mathcal{R} = 2\gamma \tilde{f}  d\ell \epsilon + \pi R_i \gamma  d\theta \epsilon
$$
By using  $d\ell = |\mathcal{R}| d\theta$, it directly leads to an expression of the radius of curvature:
\begin{equation}
\mathcal{R}= \frac{\epsilon  R_i \pi (W_e-1)}{2 \tilde{f} }  
\end{equation} 
The jet trajectory is therefore made by the repetition of arcs of a circle of constant radius, pointing alternatively upward and downward. The transition between two arcs happens when ($\vec{t}$, $\vec{e}_y$) angle equals $\pm \theta_i$. 
From geometrical considerations, the wavelength $\lambda$ of the trajectory finally writes :
\begin{equation}
\lambda=\frac{2\pi}{\tilde{f}} R_i (W_e-1) \cos(\theta_i)
\end{equation}

This expression is compared with experiments on Fig. \ref{lambda}. For a given incidence angle and different radii, the velocity dependency is tested and the best $\tilde{f}$ value is chosen to interpolate as finely as possible every data set. The best value is found to be $\tilde{f} = 0.80$ and the agreement between the experimental data and the model is rather satisfactory. This agreement is confirmed in the inset of Fig. \ref{lambda} where we plot the dimensionless wavelength $\lambda/R_i$ as a function of the Weber number and we clearly observe that all the data collapse onto the same master curve given by the model. It is worth mentioning that our parameters range holds for high Weber numbers, $We >> 1$. When $We \approx 1$, the jet  destabilizes into drops before impacting the film due to the Rayleigh-Plateau instability \cite{DeGennes2003}. It means that $\lambda$ and $\mathcal{R}$ expressions can be simplified without loss of generality by replacing $We-1\sim We$. Physically, that means that the effect described above occurs for  $|\vec{F_{1}}|  >> |\vec{F_{2}}|$. The elastic counterpart of this ``inertial-capillary'' mechanism comes consequently from the jet-film interaction and not from the curvature of the jet itself as observed for instance for meandering rivulets \cite{Drenckhan2004,LeGrand2006}.

\section{Discussion}

As discussed earlier,  the ``refraction'' is observed at high We and small incident angle. The ``absorption'' occurs for smaller We or by increasing the incident angle. 
If the jet is released and impacts the soap film with parameters corresponding to the absorption region, a transient stage characterized by a ``reflection''  on the film (i.e. bouncing of the jet on the film) is observed before absorption occurs. 
Conversely to the bouncing of drops \cite{Gilet2009}, the ``reflection'' stage is not sustainable since the air layer trapped between the jet and the film drains until it becomes too thin to prevent the coalescence of the two liquid entities. One can notice that a jet rebound on a thin liquid sheet can be sustained in the case of non-newtonian liquid exhibiting shear-thinning \cite{Versluis2006} : the so-called ``Kaye effect'' arises while  a thin layer of the liquid itself is locally sheared  at the contact zone and lubricates the latter continuously.

Finally, one can observe that the transition between the two steady regimes occurs at a Weber number largely greater than one. This is surprising since one would expect a transition around unity for phenomena balancing inertia versus capillarity. This is the case, for instance, for the impact of solid objects on a soap film \cite{LeGoff2008}. In our system the situation is different since the deformation length scale is not necessarily the same as the one of the impacting object $R_i$. As it can be seen in the inset of Fig. \ref{diagend}, the film takes the shape of a catenoid, for which the radius of curvature is largely greater than $R_i$. A quantitative study of this effect can be found in \cite{Raufaste2012}.

\section{Conclusion} 
  
To summarize, we have demonstrated the existence of three different flow classes resulting from the jet-film interaction : a refraction, an absorption and a transient reflection regime.  
The  Weber number is found to rationalize the different regimes. Models, based on momentum and mass balance equations, catch quantitatively the dependency of the different impact parameters, namely the jet  radius, velocity and incident angle. Besides its fundamental interest, this study presents a new way to guide micro-metric flows at Weber and Reynolds numbers above unity, and to probe liquid foams stability submitted to violent perturbations at the soap film scale.

\section{acknowledgments}
X. Noblin, A. Chabanov and F. Graner are thanked for the thorough reading of the manuscript.

\end{document}